\documentclass[journal,draftcls,onecolumn,12pt,twoside]{IEEEtranTCOM}
\usepackage{amsmath}
\usepackage{amssymb}
\usepackage{amsfonts}
\usepackage{amscd}
\usepackage{mathrsfs}
\usepackage{mdframed}
\usepackage{amsthm}
\usepackage{graphicx}
\usepackage{color}
\usepackage{url}
\usepackage{bm}
\usepackage{algorithm}
\usepackage{algorithmic}

\newtheorem{theorem}{Theorem}
\newtheorem{lemma}{Lemma}
\newtheorem{definition}{Definition}
\newtheorem{proposition}{Proposition}

\newtheorem{remark}{Remark}

\begin{document}

\title{Full-Diversity Space-Time Block Codes for Integer-Forcing Linear Receivers}
\author{\authorblockN{J.~Harshan, Amin Sakzad and Emanuele Viterbo}\thanks{J. Harshan is with the Department of Electrical Engineering, Indian Institute of Technology Delhi, India. Amin Sakzad is with the Clayton School of Information Technology, Monash University, Australia. Emanuele Viterbo is with the Department of Electrical and Computer Systems Engineering,
Monash University, Australia. Email: jharshan@ee.iitd.ac.in, amin.sakzad@monash.edu, emanuele.viterbo@monash.edu. Some parts of this work are published in IEEE WCNC 2017, held at San Francisco, USA.}}

\maketitle

\begin{abstract}
In multiple-input multiple-output (MIMO) fading channels, the design criterion for full-diversity space-time block codes (STBCs) is primarily determined by the decoding method at the receiver. Although constructions of STBCs have predominantly matched the maximum-likelihood (ML) decoder, design criteria and constructions of full-diversity STBCs have also been reported for low-complexity linear receivers. A new receiver architecture called Integer-Forcing (IF) linear receiver has been proposed to MIMO channels by \emph{Zhan et al.} which showed promising results for the high-rate V-BLAST encoding scheme. In this paper, we address the design of full-diversity STBCs for IF linear receivers. In particular, we are interested in characterizing the structure of STBCs that provide full-diversity with the IF receiver. Along that direction, we derive an upper bound on the probability of decoding error, and show that STBCs that satisfy the restricted non-vanishing singular value (RNVS) property provide full-diversity for the IF receiver. Furthermore, we prove that all known STBCs with the non-vanishing determinant property provide full-diversity with IF receivers, as they guarantee the RNVS property. By using the formulation of RNVS property, we also prove the existence of a full-diversity STBC outside the class of perfect STBCs, thereby adding significant insights compared to the existing works on STBCs with IF decoding. Finally, we present extensive simulation results to demonstrate that linear designs with RNVS property provide full-diversity for IF receiver. 

\end{abstract}

\begin{center}
\end{center}

\section{Introduction and Preliminaries}
\label{sec1}

Space-time coding is a powerful transmitter-side technique that assists reliable communication over multiple-input multiple-output (MIMO) fading channels. For a MIMO channel with $n_{t}$ transmit and $n_{r}$ receive antennas, a space-time block code (STBC) denoted by $\mathcal{C} \subset \mathbb{C}^{n_{t} \times T}$ is a finite set of complex matrices used to convey $\mbox{log}_{2}(|\mathcal{C}|)$ information bits to the destination \cite{TSC}. To recover the information bits all the observations collected across $n_{r}$ receive antennas and $T$ time slots at the destination are  appropriately processed by a suitable decoder $\mathcal{D}$, e.g., maximum-likelihood (ML) decoder, zero-forcing (ZF) receiver, or a minimum mean square error (MMSE) receiver. A well-known method to generate an $n_{t} \times T$ STBC is to allow the variables of an $n_{t} \times T$ linear design $\textbf{X}_{\mathcal{LD}}(x_{1}, x_{2}, \ldots, x_{K})$ take values from a finite set of complex numbers. For such STBCs, the symbol-rate is defined as $\mathcal{R} = \frac{K}{T}$ complex symbols per channel use. An STBC which is designed based on the well-known rank criterion \cite{TSC} is known to provide full-diversity for a MIMO channel, where the notion of full-diversity is as presented in Definition \ref{def:full-diversity}. However, the rank criterion applies only when the STBCs are decoded using the optimal ML decoder. 

\begin{definition}
\label{def:full-diversity}
In an $n_{t} \times n_{r}$ MIMO system with statistically independent Rayleigh fading channels between each pair of the transmitter and the receiver antennas, an STBC $\mathcal{C}$ is said to provide full-diversity for decoder $\mathcal{D}$ if at high signal-to-noise ratio (SNR), the average probability of decoder error behaves as \cite[equation (10)]{TSC}
\begin{equation*}
\mbox{Pr}(\hat{\textbf{X}} \neq \textbf{X}) \leq \frac{c}{\mbox{SNR}^{n_{t}n_{r}}},
\end{equation*}
where $\hat{\textbf{X}}$ is the decoded codeword, $\textbf{X} \in \mathcal{C}$ is the transmitted codeword, and $c$ is some constant independent of $\mbox{SNR}$.
\end{definition}

\indent Considering the high computational complexity of the ML decoder, many research groups have addressed the design and construction of full-diversity STBCs that are matched to sub-optimal \emph{linear receivers} such as the ZF and the MMSE receivers \cite{ZLW1}-\cite{FaB}. These linear receivers reduce the complexity of the decoding process by trading off some error performance with respect to ML decoder. In \cite{ShX} a new design criterion for full-diversity STBCs matched to ZF receivers is proposed, which imposes a constraint on the symbol-rate of STBCs. In particular, it has been proved that the symbol-rate of such STBCs is upper bounded by one. For some code constructions matched to ZF and MMSE linear receivers, we refer the reader to \cite{ZLW1}, \cite{SuR}, \cite{WXYL}, and \cite{FaB}. In summary, a rate loss is associated with the design of full-diversity STBCs compliant to ZF and MMSE receivers \cite{Aria13}. A comparison of the decoding complexity, diversity, and the symbol-rate of STBCs for these decoders is summarized in the first two rows of Table \ref{complexity_table}. Other than the ZF and the MMSE receivers, STBCs have also been designed for other sub-optimal decoders in MIMO channels \cite{LPS1}-\cite{ZXX}. 

\begin{table}
\begin{center}
\caption{Comparison of STBCs for various receivers: $n_{t}$ and $n_{r}$ denote the number of transmit and receive antennas, respectively.}
\begin{tabular}{|c|c|c|c|} \hline
Approach & Decoding & Spatial & Symbol-\\
& Complexity & Diversity & Rate\\
\hline
{\textbf ML} & high & $n_{t}n_{r}$ & $\leq \mbox{min}(n_{t}, n_{r})$\\
{\textbf ZF \& MMSE} & low & $n_{t}n_{r}$ & $\leq 1$\\
{\textbf IF} & low & $n_{t}n_{r}$ & $\leq \mbox{min}(n_{t}, n_{r})$\\
\hline
\end{tabular}
\end{center}
\label{complexity_table}
\end{table}
%

\begin{figure}
\begin{center}
\includegraphics[scale = 0.5]{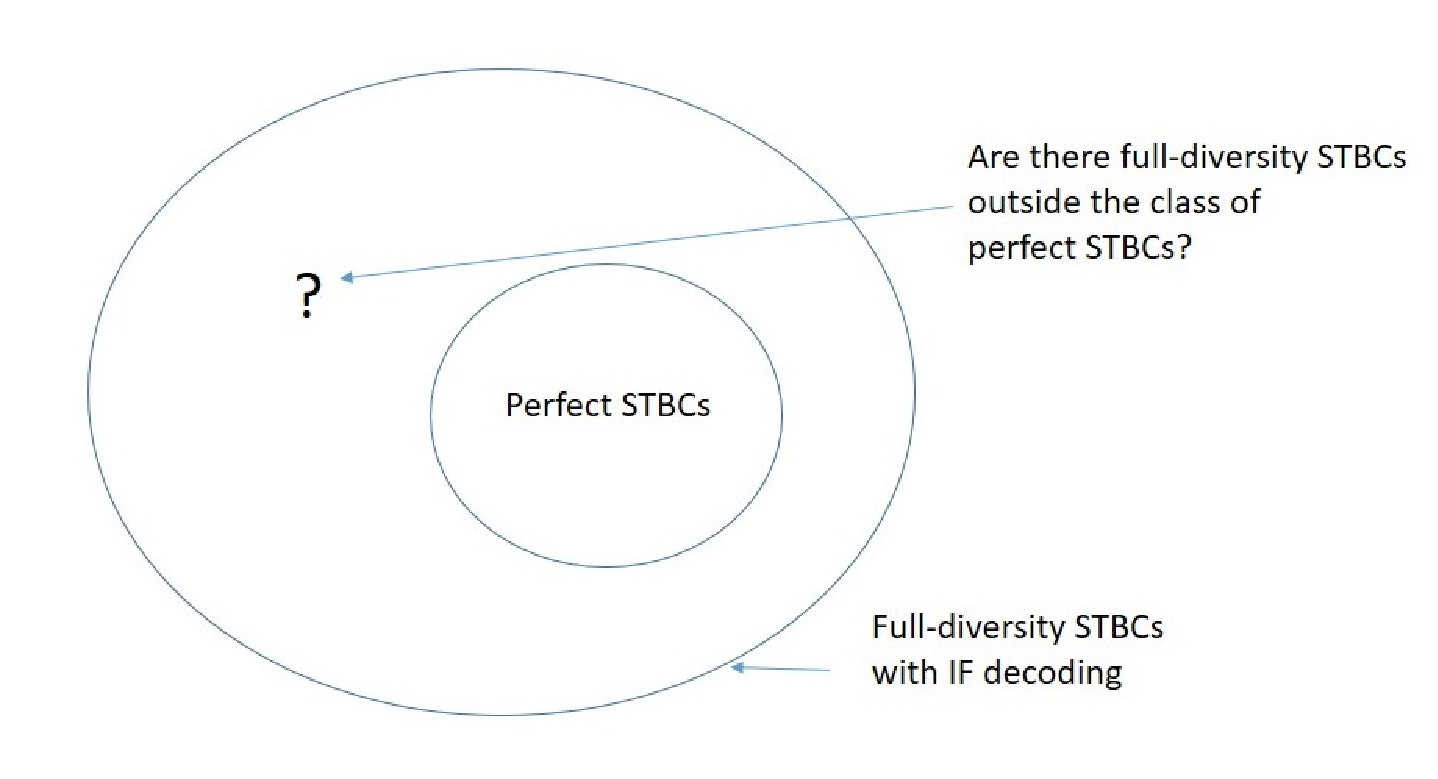}
\vspace{-1.3cm}
\end{center}
\caption{\label{fig:motivation}Depicting the motivation to develop a design criterion for STBCs amenable to integer-forcing decoding. In this work, we have answered the above question by presenting an STBC outside that class of perfect STBCs. Particularly, we have made use of the RNVS criterion to prove the full-diversity property of the proposed STBC.}
\end{figure}

\indent A new receiver architecture called integer-forcing (IF) linear receiver was recently proposed~\cite{zhan12} to attain higher rates with reduced decoding complexity. In such a framework, the source employs a layered transmission scheme and transmits independent codewords simultaneously across the layers. This has been referred to as the \emph{V-BLAST encoding scheme} in ~\cite{zhan12}. At the receiver side, each layer is allowed to decode an integer linear combination of transmitted codewords, and then recover the information by solving a system of linear equations. An outage probability based analysis has been presented to demonstrate that IF linear receivers deliver receive diversity of $n_{r}$. Although IF receivers are known to work well with the V-BLAST scheme \cite{WeC}-\cite{DoE}, not many works have investigated the suitability of IF receivers to decode STBCs. In \cite{OrE} IF receiver has been applied to decode a layered transmission scheme involving perfect STBCs. Such an architecture has been shown to achieve the capacity of any Gaussian MIMO channel up to a gap that depends on the number of transmit antennas. Although the authors of \cite{OrE} have shown that a perfect STBC provides full-diversity with IF decoding, we are not aware of other full-diversity STBCs that are outside the class of perfect STBCs. Inspired by the work in \cite{OrE}, we are interested in the broader objective of characterizing all STBCs that provide full-diversity with IF decoding. Specifically, our objective, which is as depicted in Fig. \ref{fig:motivation}, is to develop a way to search for full-diversity STBCs for IF receivers. In order to address the question in Fig. \ref{fig:motivation}, we study the error performance of IF receivers along the lines of \cite{TSC, ShX}, and propose a design criterion for constructing full-diversity STBCs. The specific contributions of this paper are summarized below:
\begin{itemize}
\item We study the application of IF linear receivers to decode STBCs in MIMO channels with Rayleigh fading characteristics. We are interested in IF linear receivers due to their ability to achieve higher rates than ZF and MMSE receivers while providing reduced decoding complexity than that of ML decoding. We present an error analysis for the IF receiver in order to obtain a design criterion for full-diversity STBCs. We first recall that the design criterion for constructing STBCs in a point-point MIMO system depends on the decoding method employed at the receiver. For instance, if the receiver chooses to employ the maximum-Likelihood (ML) decoder, then it is well known that the transmitter should employ STBCs based on the rank criterion \cite{TSC} on the code. Similarly, if the receiver chooses to employ low-complexity decoders such as MMSE, ZF, then STBCs have to be designed based on a criterion specific to the characteristics of the channel matrix \cite{ShX}. Motivated by these prior lines of work, in this work, we are interested in establishing a design criterion to construct STBCs that provide full-diversity with IF decoding. At this juncture, we would like to highlight that although the authors of [22] were the first in showing that perfect STBCs provide full-diversity with IF decoding (using the approach of diversity multiplexing trade-off), we are not aware of other full-diversity STBCs for IF decoding that are outside the class of perfect STBCs (See Fig. \ref{fig:motivation}). In our quest to characterize full-diversity STBCs for IF receivers, unlike \cite{OrE}, we do not restrict to any known classes of STBCs to start with, instead, we develop a design criterion from first principles by using an arbitrary STBC generated from a linear design.
\item One of the main contributions of this work is the formulation of the Restricted Non-Vanishing Singular value (RNVS) property on STBCs, and its subsequent connection to the average error-probability analysis of IF decoding, i.e., error-probability with not just one specific realization of the MIMO channel $\mathbf{H} \in \mathbb{C}^{n_{r} \times n_{t}}$, instead it is the error-probability averaged over several realizations of $\mathbf{H}$. Given a specific realization of the channel matrix, we first show that the error probability with IF decoding is upper bounded by a function of the minimum distance of the lattice generated by the effective channel matrix, which is a function of the space-time code and the channel matrix $\mathbf{H}$ (See \eqref{P_e_bound_2} in Lemma $1$). Subsequently, in order to obtain an upper bound on average error-probability expression, we establish a connection between the shortest vector of the lattice and the corresponding space-time codeword, which can be obtained by plugging the integer coefficients of the shortest vector into the linear design. Finally, to obtain a lower bound on the average error-probability expression, we formulate the RNVS property on the linear design, to show that those linear designs that satisfy the RNVS property will provide full-diversity when employed with IF decoding (See Theorem \ref{thm:full_diversity}). We highlight that establishing the RNVS property as a sufficient criterion for full-diversity is a novel contribution of this work, and this formulation cannot be deduced from \cite{OrE}.
\item After formulating the RNVS property, we also show that the well-known class of STBCs with the non-vanishing determinant (NVD) criterion satisfy the RNVS property, and this implies that all NVD codes provide full-diversity when decoded with IF linear receivers. Thus, we have independently confirmed the results presented in \cite{OrE} that perfect codes (which satisfy the NVD criterion \cite{Perfect1, Perfect2}) perform well with IF receivers.
\item Towards answering the question depicted in Fig. \ref{fig:motivation}, we present an example STBC that satisfies the RNVS property, but not the NVD property, thereby showcasing a full-diversity STBC outside the class of perfect STBCs (See Section \ref{subsec1_sec6}). Thus, our work adds significant insights over the existing contributions in \cite{OrE}.
\end{itemize}

The results presented in this work are significant enhancements of the work in \cite{HAV1}, wherein it was shown that non-vanishing singular value (NVS) property is a sufficient criterion for the full-diversity of STBCs. In \cite{HAV1}, although STBCs with NVS property were shown to achieve full-diversity with IF receiver, we could not prove that NVD codes also achieve full-diversity. This is because NVD property does not imply NVS property. However, in this paper, by using the RNVS framework, we have shown that all NVD codes provide full-diversity with the IF receiver. We would like to remark that the problem of constructing STBCs based on the minimum singular value criterion is not entirely new. In \cite[Ch. 9]{Tse_visw} it was shown that maximizing the minimum singular value of the difference of codeword matrices provides the \emph{approximate universality} property for STBCs in Multiple-Input Single-Output (MISO) channels, albeit with the ML decoder. Our reference to \cite{Tse_visw} serves to remind the reader about existing works which have already used minimum singular values in related applications. However, our work and \cite[Ch. 9]{Tse_visw} are fundamentally different as we address (i) IF receivers, instead of the ML decoder, (ii) MIMO environment, instead of the MISO environment, and finally, (iii) Rayleigh fading channels, instead of the approximate universality property, which caters to fading channels with arbitrary characteristics.

\indent In summary, among the class of linear receivers, we show that IF receivers can admit STBCs with larger symbol-rate than that of the MMSE and ZF receivers. As shown in Table. \ref{complexity_table}, our results highlight that unlike the traditional linear receivers such as ZF and MMSE receivers, the design criterion does not impose limitation on the symbol-rate of STBCs for IF receivers. 
\indent The rest of the paper is organized as follows: In Section \ref{sec2}, we introduce the system model of STBCs for MIMO channel, and present the decoding procedure for STBCs based on the IF receiver. In Section \ref{sec4}, we present an analysis on  error-probability with IF receiver, and propose a design criterion on STBCs that for full-diversity STBCs. In Section \ref{sec6}, we discuss the problem of constructing STBCs for IF linear receivers, and show that the existing class of STBCs based on the NVD criterio provide full-diversity with IF receivers. Finally, in Section \ref{sec7}, we present concluding remarks and some directions for future work.

\indent {\em Notations}. Boldface letters are used for vectors, and capital boldface letters for matrices. We let $\mathbb{R}$, $\mathbb{C}$, $\mathbb{Z}$, $\mathbb{Q}$, and $\mathbb{Z}[\imath]$ denote the set of real numbers, complex numbers, integers, rational numbers, and the Gaussian integers, respectively, where $\imath^2 = -1$. We let ${\textbf I}_n$ and ${\textbf 0}_n$ denote the $n\times n$ identity matrix and zero matrix and the operations $(\cdot)^T$ and $(\cdot)^H$ denote transposition and Hermitian transposition. We let $| \cdot |$ and $\| \cdot \|$ denote the absolute value of a complex number and the Euclidean norm of a vector, respectively. The operation $\mathbb{E}(\cdot)$ denotes mean of a random variable. We let $\lfloor x \rceil$ and $\lfloor {\textbf v} \rceil$ denote the closest integer to $x$ and the component-wise equivalent operation. The symbol $\textbf{X}_{j, m}$ denotes the element in the $j$-th row and $m$-th column of $\textbf{X}$. For a matrix $\textbf{X}$, the Frobenious norm $\sqrt{\sum_{j} \sum_{m}|\textbf{X}_{j, m}|^2}$ is denoted by $\|\textbf{X}\|_{F}$. The symbol $\mathcal{N}_{c}(0, 1)$ denotes circularly complex Gaussian distribution with mean zero and unit variance. For an $n_{t} \times T$ matrix $\textbf{X}$, the symbol $\sigma_{j}(\textbf{X})$ denotes the $j$-th singular value of $\textbf{X}$ for $1 \leq j \leq n_{t}$. The real and imaginary parts of a complex matrix $\textbf{X}$ is denoted by $\mbox{Re}(\textbf{X})$ and $\mbox{Im}(\textbf{X})$, respectively. The symbol $\mbox{Prob}(\cdot)$ denotes the probability operator, and the symbol $!$ represents the factorial operator. 
\section{System Model}
\label{sec2}
The $n_{t} \times n_{r}$ MIMO channel consists of a source and a destination terminal equipped with $n_{t}$ and $n_{r}$ antennas, respectively. For $1 \leq i \leq n_{t}$ and $1 \leq j \leq n_{r}$, the channel between the $i$-th transmit antenna and the $j$-th receive antenna is assumed to be flat fading and denoted by the complex number $\textbf{H}_{i, j}$. Each $\textbf{H}_{i, j}$ remains constant for a block of $T$ ($T \geq n_{t}$) complex channel uses and is assumed to take an independent realization in the next block. Statistically, we assume a Rayleigh fading channel, wherein $\textbf{H}_{i, j} \sim ~\mathcal{N}_{c}(0, 1) ~\forall i, j$ across quasi-static intervals. The source conveys information to the destination through an $n_{t} \times T$ STBC denoted by $\mathcal{C}$. We assume that a linear design
\begin{equation}
\label{linear_design}
\textbf{X}_{\mathcal{LD}}(s_{1}, \ldots, s_{2K}) = \sum_{k = 1}^{2K} \textbf{D}_{k} s_{k},
\end{equation}
in $2K$ real variables $\textbf{s} = [s_{1} ~s_{2} ~\ldots ~s_{2K}]^{T}$ is used to generate $\mathcal{C}$ by taking values from an underlying integer constellation $\mathcal{S} \subset \mathbb{Z}$. Here, the set $\{\textbf{D}_{k} \in \mathbb{C}^{n_{t} \times T}\}_{k = 1}^{2K}$ contains the weight matrices of the design. Since we use the IF linear receiver to decode the STBC, we assume that $\mathcal{S}$ is a finite ring $\mathbb{Z}_{\sqrt{M}} = \left\{0, 1, \ldots, \sqrt{M} - 1 \right\}$ for some $M$, an even power of $2$. The symbols of $\mathcal{S}$ are appropriately shifted around the origin to reduce the transmit power, and subsequently  reverted back at the receiver to retain the ring structure on $\mathcal{S}$. If $\textbf{X}(\textbf{s}) \in \mathcal{C}$ denotes a transmitted codeword matrix such that $\mathbb{E}[|\textbf{X}_{i, t}|^{2}]= 1 ~\mbox{ for } 1 \leq i \leq n_{t}, 1 \leq t \leq T$, then the received matrix $\textbf{Y} \in \mathbb{C}^{n_{r} \times n_{t}}$ at the destination is given by
\begin{equation}
\label{signal_model}
{\textbf Y} = \sqrt{\frac{P}{n_{t}}}{\textbf H}{\textbf X}(\textbf{s}) + {\textbf Z},
\end{equation}
where $\textbf{H} \in \mathbb{C}^{n_{r} \times n_{t}}$ denotes the channel matrix, $\textbf{Z} \in \mathbb{C}^{n_{r} \times T}$ denotes the additive white Gaussian noise (AWGN) with its entries that are i.i.d. as $\mathcal{N}_{c}(0, 1)$. With this, the average receive signal power-to-noise ratio (SNR) per receive antenna is $P$. Throughout the paper, we assume a coherent MIMO channel where only the receiver has the complete knowledge of $\textbf{H}$.
\begin{figure}
\begin{center}
\includegraphics[scale = 0.6]{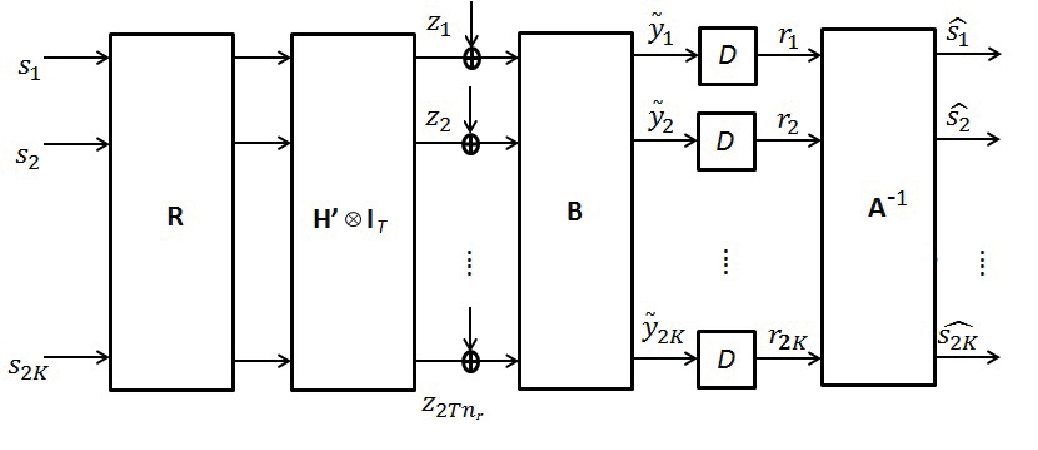}
\end{center}
\vspace{-1cm}
\caption{IF linear receiver to decode STBCs where $\textbf{R}$ is a code matrix obtained from vectorizing the components of the weight matrices $\{\textbf{D}_{k}\}_{k = 1}^{2K}$ in \eqref{linear_design}.}
\label{model1}
\end{figure}

\indent In the next subsection, we discuss the decoding procedure for STBCs based on the IF receiver.

\subsection{IF decoder for STBCs}

Since $\mathcal{C}$ is a linear dispersion code, the received matrix $\textbf{Y}$ in \eqref{signal_model} can be vectorized to obtain a noisy linear model as
\begin{eqnarray}
\label{linear_model}
\textbf{y} = \sqrt{\frac{P}{n_{t}}}\mathbf{\mathcal{H}}\textbf{s} + \textbf{z},
\end{eqnarray}
where $\mathcal{H} \in \mathbb{R}^{2n_{r}T \times 2K}$ is given by 
\begin{equation}
\label{H_dash1}
\mathcal{H} = (\textbf{H}' \otimes \textbf{I}_{T})\textbf{R},
\end{equation}
such that
\begin{equation}
\label{H_dash}
\textbf{H}' = \left[\begin{array}{rr}
\mbox{Re}(\textbf{H}) & -\mbox{Im}(\textbf{H})\\
\mbox{Im}(\textbf{H})  & \mbox{Re}(\textbf{H})\\
\end{array}\right] \in \mathbb{R}^{2n_{r} \times 2n_{t}},
\end{equation}
and $\textbf{R} \in \mathbb{R}^{2n_{t}T \times 2K}$ is a code matrix obtained from vectorizing the components of the weight matrices $\{\textbf{D}_{k}\}_{k = 1}^{2K}$. Here, the symbol $\otimes$ denotes the Kronecker product operator. After suitable scaling, \eqref{linear_model} can be equivalently written (without changing the notation) as 
\begin{eqnarray}
\label{new_linear_model}
\textbf{y} = \mathcal{H}\textbf{s} + \sqrt{\frac{n_{t}}{P}}\textbf{z}.
\end{eqnarray}
To the linear model in \eqref{new_linear_model}, we apply the IF linear receiver as shown in Fig. \ref{model1} to recover $\hat{{\textbf s}}$, a vector of decoded information symbols. For a given choice of the linear design $\mathbf{X}_{\mathcal{LD}}(\textbf{s})$ in $2K$ real variables, the number of receive antennas $n_{r}$ should satisfy the lower bound $n_{r} \geq \frac{K}{T}$ to obtain a system of linear equations in \eqref{new_linear_model} which is not information-lossy \cite{SRS_information}. As a special case, the lower bound $n_{r} \geq \frac{K}{T}$ is also required to apply ZF and MMSE decoding on \eqref{new_linear_model} since both receivers need to calculate the pseudo-inverse of $\mathcal{H}$ as their post-processing matrices. Since the post-processing matrix of IF decoding also needs pseudo-inverse of $\mathcal{H}$, the lower bound $n_{r} \geq \frac{K}{T}$ continues to be applicable with IF receivers.\footnote{Although ZF and MMSE receivers fall within the class of linear receivers, this lower bound on $n_{r}$ is not explicitly applicable in such cases since the rate (in complex symbols per channel use) of STBCs is $\frac{K}{T} \leq 1$ \cite{ShX}, and this implies that $n_{r} = 1$ suffices.} In order to decode an STBC using the IF receiver, the components of $\textbf{s}$ are restricted to take values from a subset of integers such that $\{ \mathcal{H}\textbf{s} ~|~ \textbf{s} \in \mathcal{S}^{2K}\}$ is a lattice code carved from the lattice $\Lambda = \left\lbrace \mathcal{H}\textbf{s} ~|~ \textbf{s} \in \mathbb{Z}^{2K} \right\rbrace.$ This is the reason for choosing the components of $\textbf{s}$ from the ring $\mathbb{Z}_{\sqrt{M}}$.


The goal of the IF receiver is to project $\mathcal{H}$ onto a non-singular integer matrix ${\textbf A} \in \mathbb{Z}^{2K \times 2K}$ by left multiplying $\mathcal{H}$ with a receiver filtering matrix ${\textbf B} \in \mathbb{R}^{2K \times 2n_{r}T}$. After post processing by {\textbf B}, we get
\begin{equation}~\label{eq:LRmodel}
\tilde{{\textbf y}} \triangleq {\textbf B}{\textbf y} = {\textbf B}\mathcal{H}{\textbf s}+\sqrt{\frac{n_{t}}{P}}{\textbf B}{\textbf z}.
\end{equation}
The above signal model is applicable to all linear receivers including the ZF, MMSE (both cases ${\textbf A}={\textbf I}_{2K}$), and IF (where ${\textbf A}$ is invertible over $\mathcal{S}$).
For the IF receiver formulation, we write
\begin{equation}~\label{eq:IFmodel}
\tilde{{\textbf y}}= {\textbf A}{\textbf s}+({\textbf B}\mathcal{H}-{\textbf A}){\textbf s}+\sqrt{\frac{n_{t}}{P}}{\textbf B}{\textbf z},
\end{equation}
where ${\textbf A}{\textbf s}$ is the desired signal component, and the effective noise is $({\textbf B}\mathcal{H}-{\textbf A}){\textbf s}+\sqrt{\frac{n_{t}}{P}}{\textbf B}{\textbf z}$. In particular, the effective noise power along the $m$-th row (henceforth referred to as the $m$-th layer) of $\tilde{{\textbf y}}$ for $1 \leq m \leq 2K$ is defined as
\begin{equation}~\label{quntizederrplusnoise}
g({\textbf a}_m,{\textbf b}_m)\triangleq \|{\textbf b}_m \mathcal{H}-{\textbf a}_m\|^2 \bar{E} + \frac{n_{t}}{2P}\|{\textbf b}_m\|^2,
\end{equation}
where ${\textbf a}_m$ and ${\textbf b}_m$ denote the $m$-th row of ${\textbf A}$ and $\textbf{ B}$, respectively, and $\bar{E}$ is the average energy of the constellation $\mathcal{S}$. A layer based model of the IF receiver architecture is as shown in Fig. \ref{model1}. In order to reduce the effective noise power for each layer, the term $g({\textbf a}_m,{\textbf b}_m)$ has to be minimized for each $m$ by appropriately selecting the matrices ${\textbf A}$ and ${\textbf B}$. For methods to select $\textbf{A}$ and $\textbf{B}$, we refer the reader to \cite{zhan12}, \cite{Sakzad14-1}. In order to uniquely recover the information symbols, the matrix ${\textbf A}$ must be invertible over the ring $\mathcal{S}$. In this work we are only interested in the STBC design for the IF receiver and hence, we assume that the optimal values of $\textbf{A}$ and $\textbf{B}$ are readily available.

\indent We now present a procedure for decoding STBCs using the IF linear receiver. With reference to the signal model in Section II, the decoding procedure exploits the ring structure of the constellation $\mathcal{S} = \mathbb{Z}_{\sqrt{M}} = \left\{0, 1, \ldots, \sqrt{M} - 1 \right\}$ with operations mod $\sqrt{M}$. The decoding procedure is as given below:
\begin{itemize}
\item \textbf{Step 1 (Infinite lattice decoding over $\mathbb{Z}$):} Each component of $\tilde{{\textbf y}}$ is decoded to the nearest point in $\mathbb{Z}$ to get $\hat{{\textbf y}} = \lfloor \tilde{{\textbf y}}\rceil,$ where $\lfloor \cdot \rceil$ denotes the round operation.
\item \textbf{Step 2 (Modulo operation onto $\mathcal{S}$):} Perform the modulo $\sqrt{M}$ operation on the components of $\hat{{\textbf y}}$ to obtain ${\textbf r} = \left(\hat{{\textbf y}} \mbox{ mod } \sqrt{M}\right) \in \mathcal{S}^{2K}.$
\item \textbf{Step 3 (Solving system of linear equations):} Solve the system of linear equations ${\textbf r} = {\textbf A}\hat{{\textbf s}}$ over the ring $\mathbb{Z}_{\sqrt{M}}$. If $\textbf{A}$ is invertible over the ring $\mathcal{S}$, then a unique solution is guaranteed. After solving the system of linear equations, information symbols are recovered from the components of $\hat{{\textbf s}}$. 
\end{itemize}

\indent In the above decoding procedure, \textbf{Step 2} and \textbf{Step 3} are deterministic, while \textbf{Step 1} involves recovering linear functions of the information symbols amidst noise. The critical step that reduces the complexity of the IF receiver is \textbf{Step 1}, wherein we decode the received symbol on each layer to an integer combination of the transmitted symbols, rather than jointly decoding all the symbols, which in turn would increase the decoding complexity. At each layer, the estimate of the integer linear combination can be any value in $\mathbb{Z}$. Due to this operation, some components in the vector $\mathbf{A}^{-1}\hat{\textbf{y}}$ need not lie in the base constellation $\mathcal{S}$. Thus, in \textbf{Step 2}, we have proposed a way to bring back the points in the constellation $\mathcal{S}$ through the modulo operation, by performing $\hat{\mathbf{r}} = \hat{\mathbf{y}} \mbox{ modulo } \sqrt{M}$. Although sub-optimal, the above decoding method was employed in \cite{Sakzad14-1} to show that IF decoding provides full receive-diversity in an uncoded MIMO system. Therefore, we continue to use \textbf{Step 1} to \textbf{Step 3} to derive a design criterion on full-diversity STBCs in this work. Through simulations, we will show in the later parts of the paper that implementing \textbf{Step 1 to Step 3} provides full-diversity when decoding STBCs, thereby justifying its applicability in establishing a design criterion.

In the next section, we obtain an upper bound on the probability of error for \textbf{Step 1}, and then derive a design criterion for full-diversity STBCs.


\section{Design Criterion for STBCs}
\label{sec4}

We first present an upper bound on the probability of error for \textbf{Step 1}, i.e., decoding the $m$-th layer in the infinite lattice $\mathbb{Z}$ for $1 \leq m \leq 2K$. The input to the decoder in \textbf{Step 1} is
\begin{equation*}
\tilde{{\textbf y}}_{m} = {\textbf a}_{m}{\textbf s} + ({\textbf b_{m}}\mathcal{H} - {\textbf a}_{m}){\textbf s} + \sqrt{\frac{n_{t}}{P}}\textbf{b}_{m}\textbf{z},
\end{equation*}
where $\tilde{{\textbf y}}_{m}$ denotes the $m$-th component of $\tilde{{\textbf y}}$ and $({\textbf b_{m}\mathcal{H}} - {\textbf a}_{m}){\textbf s}$ denotes the quantization noise term. 
For such a set-up, the effective noise power is given in \eqref{quntizederrplusnoise}. Note that the effective noise is not Gaussian distributed due to the quantization noise term. However, since the optimum value of $\textbf{b}_{m}$ that minimizes \eqref{quntizederrplusnoise} given $\textbf{a}_{m}$ is
\begin{equation*}
\textbf{b}_{m} = \textbf{a}_{m}\mathcal{H}^{T}\left(\frac{n_{t}}{P \bar{E}}\textbf{I}_{2n_{r}T} + \mathcal{H}\mathcal{H}^{T}\right)^{-1},
\end{equation*}
for large values of $P$, the above expression simplifies to $\textbf{b}_{m} \approx \textbf{a}_{m}\mathcal{H}^{-1},$
where $\mathcal{H}^{-1} \triangleq \left(\mathcal{H}\mathcal{H}^{T}\right)^{-1}$ denotes the pseudo-inverse of $\mathcal{H}$. While we note that $\mathcal{H}^{T}\left(\frac{n_{t}}{P \bar{E}}\textbf{I}_{2n_{r}T} + \mathcal{H}\mathcal{H}^{T}\right)^{-1}$ and $\mathcal{H}^{T}\left(\mathcal{H}\mathcal{H}^{T}\right)^{-1}$ are the post-processing matrices in the case of MMSE and ZF receivers, respectively, it is also known that the diversity performance of MMSE and ZF receivers are identical at high SNR values. Since we are interested in the performance of IF receivers for large values of $P$, we have used the zero-forcing relation between $\textbf{b}_{m}$ and $\textbf{a}_{m}$, given by $\textbf{b}_{m} = \textbf{a}_{m}\mathcal{H}^{-1}$. With this, for large values of $P$, the quantization noise term vanishes and the effective noise power is approximated by
\begin{equation*}
g({\textbf a}_m,{\textbf b}_m) = \frac{n_{t}}{2P}\|{\textbf b}_m\|^2.
\end{equation*}

\noindent Since we are interested in the full-diversity property of STBCs, which is a large SNR metric, we assume large values of $P$ in the probability of error analysis. Henceforth, we denote the probability of error for decoding the $m$-th layer in the infinite lattice  $\mathbb{Z}$ by $P_{e}(m, \mathcal{H}, \mathbb{Z})$. Using the probability of error for each layer, we now setup an upper bound on the overall probability of error for \textbf{Step 1}. We declare an error in \textbf{Step 1} if there is a decoding error in any one of the $2K$ layers. Using the union bound, the overall probability of error is bounded as
\begin{equation*}
\mbox{Pr}(\hat{{\textbf y}} \neq \textbf{As}~|~\mathcal{H}) \leq \sum_{k = 1}^{2K} P_{e}(m, \mathcal{H}, \mathbb{Z}).
\end{equation*}
After taking expectation, the average probability of error for decoding \textbf{Step 1} is
\begin{eqnarray}
\mbox{Pr}(\hat{{\textbf y}} \neq \textbf{As}) & \triangleq & \mathbb{E}_{\mathbf{H}}[\mbox{Pr}(\hat{{\textbf y}} \neq \textbf{As}~|~\mathcal{H})] \nonumber \\
& \leq & \sum_{k = 1}^{2K} \mathbb{E}_{\mathbf{H}}[P_{e}(m, \mathcal{H}, \mathbb{Z})] \nonumber \\
& = & \sum_{k = 1}^{2K} P_{e}(m, \mathbb{Z}) \label{over_all_bound},
\end{eqnarray}
where $P_{e}(m, \mathbb{Z}) \triangleq \mathbb{E}_{\mathbf{H}}[P_{e}(m, \mathcal{H}, \mathbb{Z})]$. In order to arrive at \eqref{over_all_bound}, we first obtain an upper bound on $P_{e}(m, \mathcal{H}, \mathbb{Z})$.

\begin{lemma}\textbf{(Upper Bound on Probability of Error)}
For large values of $P$, the term $P_{e}(m, \mathcal{H}, \mathbb{Z})$ is upper bounded as
\begin{equation}
\label{P_e_bound_2}
P_{e}(m, \mathcal{H}, \mathbb{Z}) \leq \mbox{exp}\left(-cP\epsilon_{1}^{2}(\Lambda)\right),
\end{equation}
where $c$ is some constant independent of $P$ and $\epsilon_{1}^{2}(\Lambda)$ is the minimum squared Euclidean distance of the lattice $\Lambda = \left\lbrace \textbf{d}\mathcal{H}^{T} ~|~ \textbf{d} \in \mathbb{Z}^{2K} \right\rbrace.$
\end{lemma}
\begin{proof}
Since the minimum Euclidean distance of $\mathbb{Z}$ is unity, an error in \textbf{Step 1} is declared if $\sqrt{\frac{n_{t}}{P}}|\textbf{b}_{m}\textbf{z}| \geq \frac{1}{2}$. Therefore, we have
\begin{eqnarray}
P_{e}(m, \mathcal{H}, \mathbb{Z}) \triangleq \mbox{Pr}\left(\sqrt{\frac{n_{t}}{P}}|\textbf{b}_{m}\textbf{z}| \geq \frac{1}{2}\right).
\end{eqnarray}
Since $\textbf{b}_{m}\textbf{z}$ is Gaussian distributed, using the Chernoff bound, $P_{e}(m, \mathcal{H}, \mathbb{Z})$ is bounded as 
\begin{eqnarray}
P_{e}(m, \mathcal{H}, \mathbb{Z}) & \leq & \mbox{exp}\left(-{\frac{P}{4n_{t}\|{\textbf b}_m\|^2}}\right) \nonumber\\
& = & \mbox{exp}\left(-{\frac{P}{4n_{t}\|{\textbf a}_m \mathcal{H}^{-1}\|^2}}\right). \label{P_e_bound1}
\end{eqnarray}
If $\textbf{a}_{m}$ and $\textbf{b}_{m}$ are chosen appropriately as in \cite{zhan12}, then the upper bound 
\begin{equation}
\label{bound1}
\|{\textbf a}_m \mathcal{H}^{-1}\|^2 \leq \epsilon_{2K}^{2}(\Lambda^{*})
\end{equation}
holds good where $\epsilon_{2K}^{2}(\Lambda^{*})$ denotes the $2K$-th successive minimum of the dual lattice
\begin{equation*}
\Lambda^{*} = \left\lbrace \textbf{d}\mathcal{H}^{-1} ~|~ \forall \textbf{d} \in \mathbb{Z}^{2K} \right\rbrace.
\end{equation*}
Here $\mathcal{H}^{-1}$ is a generator of the dual lattice $\Lambda^{*}$ of the lattice given by $\Lambda = \left\lbrace \textbf{d}\mathcal{H}^{T} ~|~ \forall \textbf{d} \in \mathbb{Z}^{2K} \right\rbrace,$ which is generated by the rows of $\mathcal{H}^{T}$. Thus we have the relation (see Lemma $4$ in \cite{zhan12})
\begin{equation}
\label{bound2}
\epsilon_{2K}^{2}(\Lambda^{*}) \leq \frac{2K^{3} + 3K^{2}}{\epsilon_{1}^{2}(\Lambda)},
\end{equation}
where $\epsilon_{1}^{2}(\Lambda)$ is the minimum squared Euclidean distance of the lattice $\Lambda$. Using the upper bounds of \eqref{bound1} and \eqref{bound2} in \eqref{P_e_bound1}, the  probability of error for decoding the $m$-th layer is upper bounded as
\begin{equation}
\label{P_e_bound_2}
P_{e}(m, \mathcal{H}, \mathbb{Z}) \leq \mbox{exp}\left(-cP\epsilon_{1}^{2}(\Lambda)\right),
\end{equation}
where $c = \frac{1}{4n_{t}(2K^{3} + 3K^{2})}$ is a constant. This completes the proof.
\end{proof}

The error probability expression in \eqref{P_e_bound_2} is for a specific realization of $\mathcal{H}$. We now take the average of \eqref{P_e_bound_2} over different channel realizations. Let $\xi$ be the random variable used to represent $\epsilon_{1}^{2}(\Lambda)$, which is a function of $\mathbf{H}$, and let $\epsilon$ denote a realization of $\xi$. Taking expectation of \eqref{P_e_bound_2} over $\xi$, and denoting $\mathbb{E}_{\xi}[P_{e}(m, \mathcal{H}, \mathbb{Z})]$ by $P_{e}(m, \mathbb{Z})$, we get
\begin{equation}
\label{bounding_second_term_prev}
P_{e}(m, \mathbb{Z}) = \int_{\epsilon = 0}^{1} P_{\xi}(\epsilon)\mbox{exp}\left(-cP\epsilon\right)d\epsilon + \int_{\epsilon = 1}^{\infty} P_{\xi}(\epsilon)\mbox{exp}\left(-cP\epsilon\right) d\epsilon,
\end{equation} 
where $P_{\xi}(\epsilon)$ is the probability density function of the random variable $\xi$. Using the lower bound $\epsilon \geq 1$ in the second term of \eqref{bounding_second_term_prev}, we have 
\begin{equation*}
\int_{\epsilon = 1}^{\infty} P_{\xi}(\epsilon)\mbox{exp}\left(-cP\epsilon\right) d\epsilon < \int_{\epsilon = 1}^{\infty} P_{\xi}(\epsilon)\mbox{exp}\left(-cP\right) d\epsilon = \mbox{exp}\left(-cP\right) \int_{\epsilon = 1}^{\infty} P_{\xi}(\epsilon) d\epsilon < \mbox{exp}\left(-cP\right).
\end{equation*}
Therefore, the expression in \eqref{bounding_second_term_prev} can be upper bounded as   
\begin{equation}
\label{bounding_second_term}
P_{e}(m, \mathbb{Z}) < \int_{\epsilon = 0}^{1} P_{\xi}(\epsilon)\mbox{exp}\left(-cP\epsilon\right)d\epsilon + \mbox{exp}\left(-cP\right).
\end{equation} 
Since the second term experiences exponential fall as a function of $P$, henceforth, we focus on the dominant term, which is the first term of the above expression. 

\subsection{Connection to STBCs}

To establish a relation between the upper bound in $\eqref{bounding_second_term}$ and the structure of linear designs, let us first understand the structure of integer vectors $\mathbf{d} \in \mathbb{Z}^{2K}$ that result in $\xi$ less than one. 
\begin{definition}
Let us define the set $\mathcal{D}$ given by $$\mathcal{D} \triangleq \{\mathbf{d} \in \mathbb{Z}^{2K} ~|~ \xi = \|{\mathbf d} \mathcal{H}^{T}\|^2 < 1 \mbox{~for~some~} \mathbf{H} \in \mathbb{C}^{n_{r} \times n_{t}}\},$$ which contains $\mathbf{d} \in \mathbb{Z}^{2K}$ that result in $\xi$ less than one. 
\end{definition}
Since the components of $\mathbf{H}$ are distributed as $\mathcal{N}_{c}(0, 1)$ and $\mathbf{R}$ is a constant matrix, using \eqref{H_dash1} it is straightforward to show that the set $\mathcal{D}$ is non-empty. We next show that the elements of $\mathcal{D}$ are bounded within a sphere of finite radius with high probability, i.e., for $\mathbf{d} \in \mathcal{D}$, we show that $Prob(\|\mathbf{d}\|^2 \leq C)$, for some large constant $C$, is close to one. Before stating such a result, we recall some results from random matrix theory~\cite{Khatri}-\cite{Burel} and linear algebra~\cite{HLA}.
\begin{lemma}
The joint probability density function (PDF) of the eigenvalues of the unordered, central, uncorrelated Wishart matrix ${\bf W}={\bf H}^H{\bf H}$ can be written as
$$f_{\bf \lambda}(x_{1}, x_{2}, \ldots, x_{n_{t}}) = c_2 \left(\det\left({\bf V}(x_{1}, x_{2}, \ldots, x_{n_{t}})\right)\right)^2\prod_{\ell=1}^{n_t}\exp(-x_\ell),$$
where $x_{l} = \lambda_{l}(\mathbf{W})$ is the $l$-th eigenvalue  of $\mathbf{W}$, the $n_{t} \times n_{t}$ matrix ${\bf V}({\bf x})$ denotes the Vandermonde matrix whose $(i,j)$-th component is given by $x_j^{i-1}$, and finally the constant $c_{2}$ is a normalizing factor.
\end{lemma}
After marginalizing the higher-order eigenvalues, it is straightforward to obtain the following complimentary cumulative distribution function (CCDF) on the smallest eigenvalue value of the Wishart matrix~\cite{Khatri}-\cite{Burel}.
\begin{lemma}~\label{lem:randmat}
The CCDF of the least eigenvalue of the $n_{t} \times n_{t}$ Wishart matrix ${\bf W}$ is:
$$Prob\left(\lambda_{\min}({\bf W})>c_1\right) = c_2(n_t !)|\det({\bf M}_{c_1})|,$$
where ${\bf M}_{c_1}$ is an $n_t \times n_t$ matrix with $(i,j)$-th entry being
the tail of Gamma function of order $i+j-1$, i.e., ${\bf M}_{c_1}(i, j) = \int_{c_1}^\infty w^{i+j-1}\exp(-w)dw$. Furthermore, since $Prob(\lambda_{\min}({\bf W})\geq0) = 1$ the normalizing constant is $c_2 = \left(n_t ! |\det({\bf M}_0)|\right)^{-1}$. 
\end{lemma}

Using the above standard results from matrix theory, we show that the elements of $\mathcal{D}$ are bounded within a sphere of finite radius with high probability.

\begin{proposition}
\label{bounding_prop}
With $c_3$ denoting the least singular value of the constant matrix ${\bf R}$, we have 
\begin{equation*}
Prob\left(\|\mathbf{d}\|^2 \leq \frac{1}{c^2_3c_1}\right) = \frac{|\det({\bf M}_{c_1})|}{|\det({\bf M}_0)|}.
\end{equation*}
\end{proposition}
\begin{IEEEproof}
From the definition of the set $\mathcal{D}$, its member ${\bf d} \in \mathcal{D}$ is such that $\|{\bf d}\mathcal{H}^T\|^2 < 1$, for some $\mathcal{H}$. On the other hand, $\|{\bf d}\mathcal{H}^T\|^2$ can be lower bounded as follows:
\begin{eqnarray}
\label{ineq1}
\|{\bf d}\mathcal{H}^T\|&\geq&\|{\bf d}\|\sigma_{\min}\left(\mathcal{H}^T\right),\\
\label{ineq2}
&=&\|{\bf d}\|\sigma_{\min}\left(\mathcal{H}\right),\\
\label{ineq3}
&=&\|{\bf d}\|\sigma_{\min}\left(({\bf H}'\otimes {\bf I}){\bf R}\right),\\
\label{ineq4}
&\geq&\|{\bf d}\|\sigma_{\min}\left(({\bf H}'\otimes {\bf I})\right)\sigma_{\min}\left({\bf R}\right),\\
\label{ineq5}
&\geq & c_3\|{\bf d}\|\sigma_{\min}({\bf H}),
\end{eqnarray}
where the relations in \eqref{ineq1}-\eqref{ineq5} follow from basic results in matrix theory. In \eqref{ineq5}, the constant $c_{3}$ denotes the constant $\sigma_{\min}\left({\bf R}\right)$. Using the definition of $\mathcal{D}$ on \eqref{ineq5}, we have $c_3\|{\bf d}\|\sigma_{\min}({\bf H}) < 1$, which implies that $\|{\bf d}\| < \frac{1}{c_{3}\sigma_{\min}({\bf H})}$. Finally, from Lemma ~\ref{lem:randmat}, since $\sigma^{2}_{\min}({\bf H}) > c_{1}$ with probability $|\det({\bf M}_{c_1})|/|\det({\bf M}_0)|$, we have $\|{\bf d}\|^2 < \frac{1}{c^2_{3}c_{1}}$ with the same probability.
\end{IEEEproof}

\begin{figure}
\begin{center}
\includegraphics[scale = 0.4]{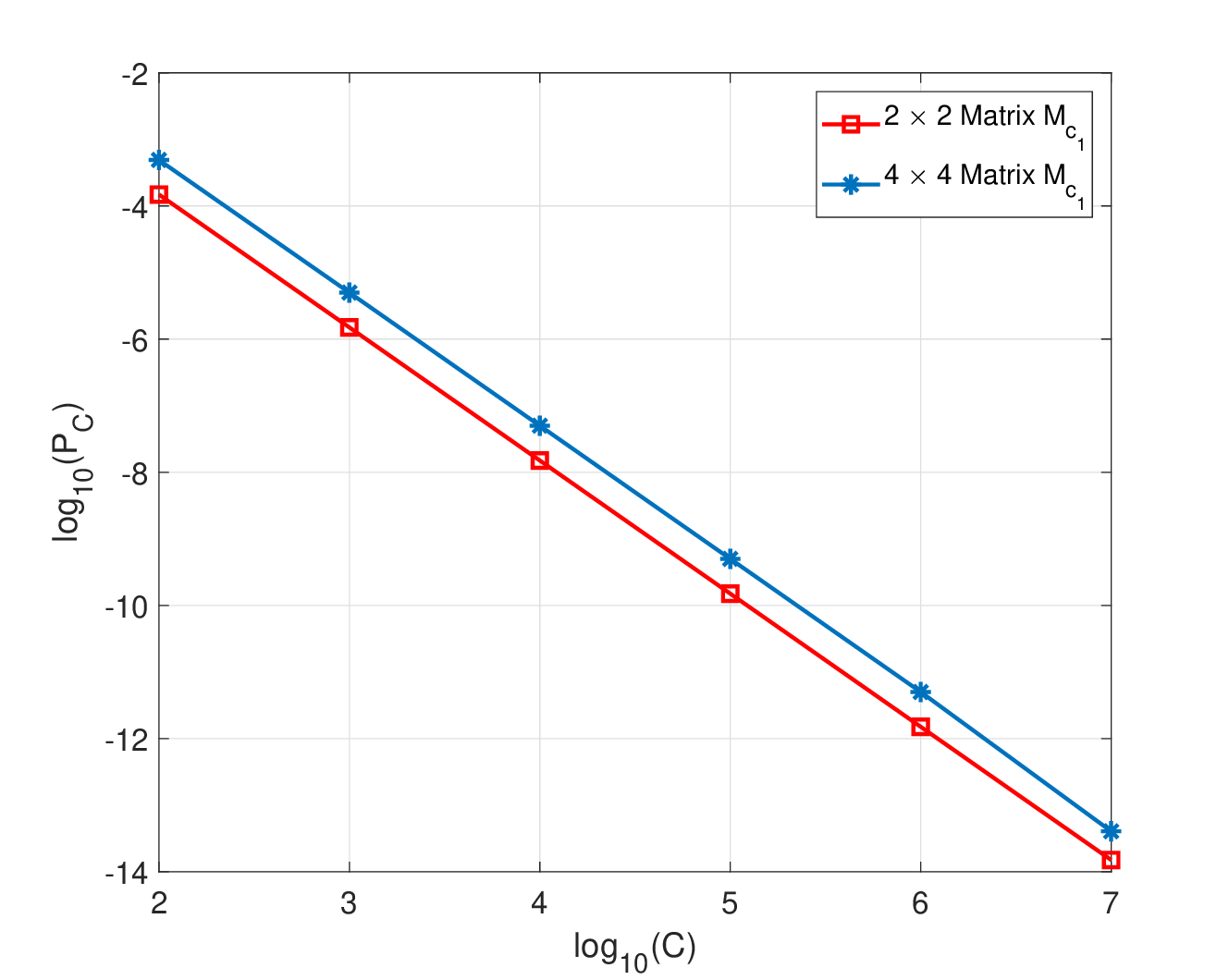}
\end{center}
\vspace{-1cm}
\caption{\label{fig:C_choice}Plots depicting the choice of $C$ for a given value of $Prob(\|\bar{\mathbf{d}}\|^{2} > C)$. Since we are looking at negligible values of $P_{C} \triangleq Prob(\|\bar{\mathbf{d}}\|^{2} > C)$ and large values of $C$, we have used $log_{10}$ scale to capture the exponents.}
\end{figure}

The above proposition shows that with small values of $c_{1}$, the elements of the set $\mathcal{D}$ are bounded within a sphere of finite radius with high probability. Recall that $\epsilon_{1}^{2}(\Lambda) = \|\bar{{\mathbf d}} \mathcal{H}^{T}\|^2,$ where $\bar{{\mathbf d}} = arg \min_{\mathbf{d} \in \mathbb{Z}^{2K}} \|{\mathbf d} \mathcal{H}^{T}\|^2$. Using this representation, we define two sets $\mathcal{E}_{C}$ and $\mathcal{E}_{\bar{C}}$ as
\begin{equation}
\label{eq:E_set}
\mathcal{E}_{C} = \{\epsilon_{1}^{2}(\Lambda) < 1 ~|~ \|\bar{{\mathbf d}}\|^2 \leq C\} \mbox{ and } \mathcal{E}_{\bar{C}} = \{\epsilon_{1}^{2}(\Lambda) < 1 ~|~ \|\bar{{\mathbf d}}\|^2 > C\},
\end{equation}
where $C >> 0$. Using $\mathcal{E}_{C}$ and $\mathcal{E}_{\bar{C}}$, we rewrite \eqref{bounding_second_term} as
\begin{equation}
\label{final_approx_prev}
P_{e}(m, \mathbb{Z}) < \int_{\epsilon \in \mathcal{E}_{C}} P_{\xi}(\epsilon)\mbox{exp}\left(-cP\epsilon\right)d \epsilon + \int_{\epsilon \in \mathcal{E}_{\bar{C}}} P_{\xi}(\epsilon)\mbox{exp}\left(-cP\epsilon\right)d \epsilon + \mbox{exp}\left(-cP\right).
\end{equation} 
We can further upper bound the above expression as
\begin{eqnarray}
\label{final_approx}
P_{e}(m, \mathbb{Z}) < \int_{\epsilon \in \mathcal{E}_{C}} P_{\xi}(\epsilon)\mbox{exp}\left(-cP\epsilon\right)d \epsilon + Prob(\|\bar{\mathbf{d}}\|^{2} > C) + \mbox{exp}\left(-cP\right),
\end{eqnarray} 
wherein we use the relation $$\int_{\epsilon \in \mathcal{E}_{\bar{C}}} P_{\xi}(\epsilon)\mbox{exp}\left(-cP\epsilon\right)d \epsilon < \int_{\epsilon \in \mathcal{E}_{\bar{C}}} P_{\xi}(\epsilon)d \epsilon = Prob(\epsilon \in \mathcal{E}_{\bar{C}}) = Prob(\|\bar{\mathbf{d}}\|^{2} > C),$$
to obtain \eqref{final_approx} from \eqref{final_approx_prev}. From Proposition \ref{bounding_prop}, we can choose a sufficiently large $C$ such that $Prob(\|\bar{\mathbf{d}}\|^{2} > C)$ is negligible. In particular, the exact relation between $C$ and $Prob(\|\mathbf{d}\|^{2} > C)$ can be obtained from Proposition \ref{bounding_prop} as follows. From Proposition \ref{bounding_prop}, the term $c_{3} > 0$ is a constant (as it is a function of the code matrix $\mathbf{R}$), whereas the other term $c_{1} > 0$ can be varied to drive the ratio $\frac{\mbox{det}(\mathbf{M}_{c_{1}})}{\mbox{det}(\mathbf{M}_{0})}$ close to $1$. Note that the entries of the matrices $\mathbf{M}_{c_{1}}$ and $\mathbf{M}_{0}$ can be computed using incomplete Gamma functions as given in Lemma \ref{lem:randmat}. If we would like to fix $Prob(\|\bar{\mathbf{d}}\|^{2} > C) < \epsilon$ for some small $\epsilon > 0$ of our choice, then we can choose $c_{1}$ sufficiently small to drive $\frac{det(\mathbf{M}_{c_{1}})}{det(\mathbf{M}_{0})} \geq 1 - \epsilon$. The corresponding value of $C$ is $\frac{1}{c_{3}c_{1}}$. To exemplify the right choice of $C$, we have used incomplete Gamma functions given in Lemma \ref{lem:randmat} to plot $\mbox{log}_{10}(P_{C})$ as a function of $\mbox{log}_{10}(C)$ in Fig. \ref{fig:C_choice}, where $P_{C} \triangleq Prob(\|\bar{\mathbf{d}}\|^{2} > C)$. The plots in figure show that we can choose a sufficiently large value of $C$ to neglect the second term in \eqref{final_approx}. Henceforth, we only consider the first term of \eqref{final_approx}. In order to analyze the dominant term, we introduce the following STBCs. 

An STBC $\mathcal{C}_{\mathcal{D}}$ generated from a linear design $\textbf{X}_{\mathcal{LD}}$ and $\mathcal{D}$ is given by
\begin{equation}
\mathcal{C}_{\mathcal{D}} \triangleq \left\lbrace \textbf{X} = \sum_{k = 1}^{2K} \textbf{D}_{k} s_{k} ~|~ \mathbf{s} \in \mathcal{D} \right\rbrace. 
\end{equation}

\begin{definition}
Let $\mathcal{B}_{C} = \{\mathbf{r} \in \mathbb{Z}^{2K} ~|~ \|\mathbf{r}\|^{2} \leq C\}$ denote the set of all integer vectors bounded within a sphere of radius $C$. Using $\mathcal{D}$ and $\mathcal{B}_{C}$, we obtain a subset of $\mathcal{D}$ defined by $\mathcal{D}_{C} \triangleq \mathcal{D} \cap \mathcal{B}_{C}.$
\end{definition}

Similar to $\mathcal{C}_{\mathcal{D}}$, an STBC $\mathcal{C}_{\mathcal{D}_{C}}$ generated from $\mathcal{D}_{C}$ is given by
\begin{equation}
\label{restricted_code}
\mathcal{C}_{\mathcal{D}_{C}} \triangleq \left\lbrace \textbf{X} = \sum_{k = 1}^{2K} \textbf{D}_{k} s_{k} ~|~ \mathbf{s} \in \mathcal{D}_{C} \right\rbrace. 
\end{equation}
Since $\mathcal{D}_{C}$ is bounded within a circle of finite radius, $\mathcal{C}_{\mathcal{D}_{C}}$ is also finite in size. 
We let $\sigma_{min}(\textbf{X}) = \min_{1 \leq j \leq n_{t}} \sigma_{j}(\textbf{X})$ denote the minimum singular value of $\textbf{X}$. With that, the minimum singular value of $\mathcal{C}_{\mathcal{D}_{C}}$ is given by
\begin{equation*}
\sigma_{min}(\mathcal{C}_{\mathcal{D}_{C}}) \triangleq \min_{\textbf{X} \in \mathcal{C}_{\mathcal{D}_{C}}, \textbf{X} \neq \textbf{0}} \sigma_{min}(\textbf{X}).
\end{equation*}
Using the above definition of the minimum singular value of the code $\mathcal{C}_{\mathcal{D}_{C}}$, we define a special class of linear designs as follows:
\begin{definition}\textbf{(Restricted non-vanishing singular value property)}
A linear design $\textbf{X}_{\mathcal{LD}}$ is said to have the restricted non-vanishing singular value (RNVS) property over $\mathbb{Z}$ if the corresponding STBC $\mathcal{C}_{\mathcal{D}_{C}}$ in \eqref{restricted_code} satisfies $\sigma_{min}(\mathcal{C}_{\mathcal{D}_{C}}) \neq 0.$
\end{definition}
\indent We now connect the RNVS property of $\textbf{X}_{\mathcal{LD}}$ and the full-diversity property of $\mathcal{C}$ for the IF receiver in the following theorem.
\begin{theorem}\label{thm:full_diversity}\textbf{(Full-Diversity Design Criterion)}
If the linear design $\textbf{X}_{\mathcal{LD}}$ has the RNVS property, then any STBC $\mathcal{C}$ generated from $\textbf{X}_{\mathcal{LD}}$ provides full-diversity with the IF linear receiver.
\end{theorem}
\begin{proof}
The minimum squared distance $\epsilon_{1}^{2}(\Lambda)$ is a random variable as it is a function of the channel $\mathbf{H}$. Specifically, we write $\epsilon_{1}^{2}(\Lambda)$ as
\begin{equation*}
\epsilon_{1}^{2}(\Lambda) = \|\bar{{\textbf d}} \mathcal{H}^{T}\|^2 = \|\mathcal{H}\bar{{\textbf d}}^{T}\|^2
\end{equation*}
for $\bar{{\textbf d}} = arg \min_{\mathbf{d} \in \mathbb{Z}^{2K}} \|{\textbf d} \mathcal{H}^{T}\|^2$. Further, $\epsilon_{1}^{2}(\Lambda)$ can be written as
\begin{eqnarray*}
\epsilon_{1}^{2}(\Lambda) & = & \|\textbf{H}\textbf{X}\|_{F}^2 = \mbox{Trace} \left(\textbf{H}\textbf{U}\Sigma\textbf{U}^{H}\textbf{H}^{H}\right),
\end{eqnarray*}
where $\textbf{X} \in \mathcal{C}_{\Lambda}$ is obtained by using $\mathbf{s} = \bar{{\textbf d}}$, $\textbf{U}\Sigma\textbf{U}^{H}$ is a singular value decomposition of $\textbf{X}\textbf{X}^{H}$, and $\Sigma$ is the diagonal matrix comprising of the square of the singular values $\sigma_{j}(\textbf{X})$ for $1 \leq j \leq n_{t}$. By denoting $\textbf{HU} = \textbf{G}$, we write
\begin{equation*}
 \epsilon_{1}^{2}(\Lambda) = \mbox{Trace} \left(\textbf{G}\Sigma\textbf{G}^{H}\right)
 = \sum_{j = 1}^{n_{t}} \|\textbf{g}_{j}\|^2 \sigma^{2}_{j}(\textbf{X}),
\end{equation*}
where $\textbf{g}_{j}$ is the $j$-th column of $\textbf{G}$ and $\sigma_{j}(\textbf{X})$ denotes the $j$-th singular value of $\textbf{X}$, which is a function of $\bar{{\textbf d}}$, which in turn is a function of the channel $\textbf{H}$. If the STBC has the RNVS property, then for any $\bar{\textbf{d}} \in \mathcal{D}_{C}$ we apply $\sigma^{2}_{j}(\textbf{X}) \geq \sigma^{2}_{min}(\mathcal{C}_{\mathcal{D}_{C}}) ~\forall j,$ and hence,
\begin{eqnarray*}
\epsilon_{1}^{2}(\Lambda) & \geq & \sum_{j = 1}^{n_{t}} \|\textbf{g}_{j}\|^2 \sigma^{2}_{min}(\mathcal{C}_{\mathcal{D}_C}),
\end{eqnarray*}
Plugging the above lower bound in \eqref{final_approx}, we get
\begin{eqnarray}
P_{e}(m, \mathbb{Z}) & < & \int_{\epsilon \in \mathcal{E}_{C}} P_{\xi}(\epsilon)\mbox{exp}\left(-cP\sum_{j = 1}^{n_{t}} \|\textbf{g}_{j}\|^2 \sigma^{2}_{min}(\mathcal{C}_{\mathcal{D}_{C}})\right) d \epsilon \nonumber \\ 
& & + Prob(\|\bar{\mathbf{d}}\|^{2} > C) + \mbox{exp}\left(-cP\right). \nonumber\\
 & < & \int_{\epsilon \geq 0} P_{\xi}(\epsilon)\mbox{exp}\left(-cP\sum_{j = 1}^{n_{t}} \|\textbf{g}_{j}\|^2 \sigma^{2}_{min}(\mathcal{C}_{\mathcal{D}_{C}})\right) d \epsilon \label{range_extension} \\
& & + Prob(\|\bar{\mathbf{d}}\|^{2} > C) + \mbox{exp}\left(-cP\right) \nonumber\\
 & = & \mathbb{E}_{\xi} \left[\mbox{exp}\left(-cP\sum_{j = 1}^{n_{t}} \|\textbf{g}_{j}\|^2 \sigma^{2}_{min}(\mathcal{C}_{\mathcal{D}_{C}})\right)\right] + Prob(\|\bar{\mathbf{d}}\|^{2} > C) + \mbox{exp}\left(-cP\right) \nonumber \\
  & = & \mathbb{E}_{\mathbf{G}} \left[\mbox{exp}\left(-cP\sum_{j = 1}^{n_{t}} \|\textbf{g}_{j}\|^2 \sigma^{2}_{min}(\mathcal{C}_{\mathcal{D}_{C}})\right)\right] + Prob(\|\bar{\mathbf{d}}\|^{2} > C) + \mbox{exp}\left(-cP\right)
  \label{overall_expression}.
\end{eqnarray}
Notice that the second inequality in \eqref{range_extension} is obtained by extending the range of $\|\bar{\mathbf{d}}\|^{2}$ from $C$ to $\infty$. Since $\textbf{U}$ is a unitary matrix, the distribution of $\textbf{G}$ is same as that of $\textbf{H}$. Also, as $\sigma^{2}_{min}(\mathcal{C}_{\mathcal{D}_{C}})$ is a constant and independent of $\mathbf{H}$, the random variables in the exponent are $\{ \|\textbf{g}_{j}\|^2, 1 \leq j \leq n_{t} \}$, which are chi-square distributed with degrees of freedom $2n_{r}$. By averaging the first term of \eqref{overall_expression} over different realizations of $\|\textbf{g}_{j}\|^2$, we obtain
\begin{eqnarray*}
P_{e}(m, \mathbb{Z}) & < & \left(\frac{1}{1 + cP\sigma^{2}_{min}(\mathcal{C}_{\mathcal{D}_{C}})}\right)^{n_{t}n_{r}} + Prob(\|\bar{\mathbf{d}}\|^{2} > C) + \mbox{exp}\left(-cP\right).
\end{eqnarray*}
Since $P$ is dominant and $\sigma_{min}(\mathcal{C}_{\mathcal{D}_{C}}) \neq 0$, $P_{e}(m, \mathbb{Z})$ is upper bounded as
\begin{eqnarray*}
P_{e}(m, \mathbb{Z}) & < & \left(\frac{1}{cP \sigma^{2}_{min}(\mathcal{C}_{\mathcal{D}_{C}})}\right)^{n_{r} n_{t}} + Prob(\|\bar{\mathbf{d}}\|^{2} > C) + \mbox{exp}\left(-cP\right).
\end{eqnarray*}
Notice that the above upper bound is only a function of $\mathcal{C}_{\mathcal{D}_{C}}$, and it is independent of the constellation $\mathcal{S}$. This shows that any STBC carved from a linear design with the RNVS property provides diversity of $n_{t}n_{r}$ independent of the size of $\mathcal{S}$.
\end{proof}

\begin{remark}
In contrast to the criterion in \cite{HAV1}, where the NVS property is applicable on an infinite STBC, the proposed RNVS property is a relaxed criterion applicable on a finite STBC $\mathcal{C}_{\mathcal{D}_{C}}$, where the constant $C$ is chosen sufficiently large such that $Prob(\|\bar{\mathbf{d}}\|^{2} > C) < \delta$ for some $\delta > 0$ of our choice.
\end{remark}

\section{Full-Diversity STBCs for IF receiver}
\label{sec6}
In the previous section, we have shown that linear designs with the RNVS property can generate full-diversity STBCs for IF receiver. In the STBC literature, there is a special class of linear designs that has a similar property called the non-vanishing determinant (NVD) property. For a linear design, such a property holds when $$\inf_{\textbf{X} \in \mathcal{C}_{\infty}, \textbf{X} \neq \textbf{0}} \mbox{det}(\textbf{X}\textbf{X}^{H}) \neq 0,$$ where
\begin{equation*}
\mathcal{C}_{\infty} \triangleq \left\lbrace \textbf{X} = \sum_{k = 1}^{2K} \textbf{D}_{k} s_{k} ~|~ \mathbf{s} \in \mathbb{Z}^{2K} \right\rbrace. 
\end{equation*}
Using the relation between $\mbox{det}(\textbf{X}\textbf{X}^{H})$ and $\sigma_{min}(\textbf{X})$, in the following proposition, we show that the NVD property implies the RNVS property. 
\begin{proposition}
\label{prop:nvd_nvs}
A linear design $\textbf{X}_{\mathcal{LD}}$ satisfies the RNVS property if it satisfies the NVD property.
\end{proposition}
\begin{proof}
Let us start with a linear design that satisfies the NVD property. Since $\mathcal{C}_{\mathcal{D}_{C}} \subset \mathcal{C}_{\infty}$, it follows that $$\min_{\textbf{X} \in \mathcal{C}_{\Lambda, C}, \textbf{X} \neq \textbf{0}} \mbox{det}(\textbf{X}\textbf{X}^{H}) \neq 0.$$ Furthermore, since $\sigma^{2}_{1}(\textbf{X})\sigma^{2}_{2}(\textbf{X})\ldots \sigma^{2}_{n_{t}}(\textbf{X}) = \mbox{det}(\textbf{X}\textbf{X}^{H})$, we have $\sigma_{min}(\mathcal{C}_{\mathcal{D}_{C}}) \neq 0$.
Thus the NVD property implies the RNVS property.
\end{proof}
The above proposition proves that linear designs with the NVD property provide full-diversity STBCs for the IF receiver. With $n_{t} = 2$, some well known examples for NVD designs include the Golden code and the Silver code designs, which carry $K = 4$ complex symbols over $T = 2$ channel uses. This result implies that the Golden code and the Silver code designs are amenable to IF decoding in a $2 \times n_{r}$ MIMO system as long as $n_{r} \geq 2$. If $n_{r} =1$, the effective channel matrix $\mathcal{H}$ results in an underdetermined system of linear equations given in \eqref{linear_model}, and therefore IF decoding is no longer applicable. In general, the applicability of NVD property along with the lower-bound $n_{r} \geq \frac{K}{T}$ implies that high-rate STBCs are amenable to IF decoding as long as the number of receive antennas is sufficiently large to implement the IF decoder.

\begin{figure}
\begin{center}
\includegraphics[scale = 0.6]{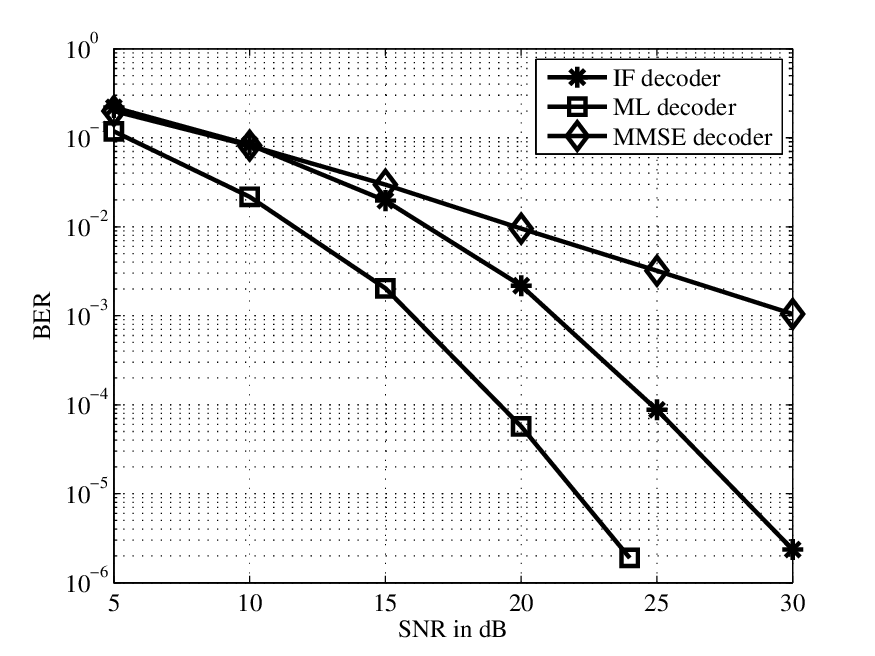}
\vspace{-0.5cm}
\end{center}
\caption{BER comparison of Golden code with IF linear receiver, MMSE receiver, and the ML decoder.}
\label{ber1}
\end{figure}

\begin{figure}
\begin{center}
\includegraphics[scale = 0.6]{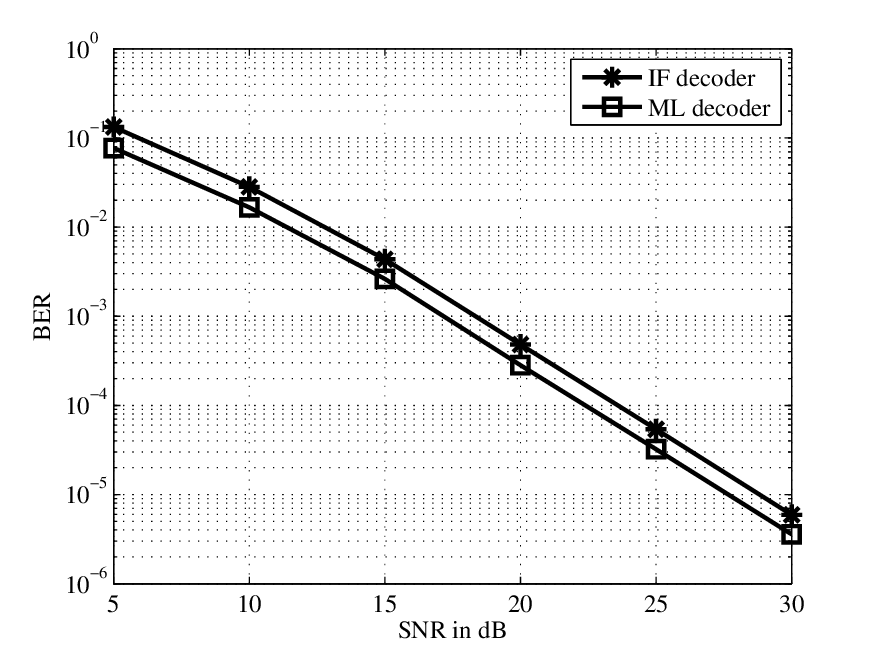}
\end{center}
\vspace{-0.5cm}
\caption{BER comparison of Alamouti code with IF linear receiver and the ML decoder.}
\label{berl}
\end{figure}

In Fig. \ref{ber1}, we present the bit error rate (BER) of the Golden code \cite{BRV} for the $2 \times 2$ MIMO channel when decoded with (\emph{i}) the IF receiver, (\emph{ii}) the ML decoder (realized using the sphere decoder), and (\emph{iii}) the MMSE decoder. The plots confirm that the Golden code provides full-diversity with the IF receiver as the BER curve of the IF receiver falls parallel to that of the ML decoder. For the simulation results, the lattice-reduction method proposed in \cite{Sakzad14-1} is used throughout the paper to compute the $\textbf{A}$ and $\textbf{B}$ matrices for the IF receiver. Similarly, we use the Alamouti design given by 
\begin{equation}
\label{example_code_with_nvs}
\textbf{X}_{A} = \left[\begin{array}{cc}
x_{1} & x_{2}\\
-x^{*}_{2} & x^{*}_{1}\\
\end{array}\right],
\end{equation}
to showcase the results with the IF receiver. Using the structure of the above design, it can be shown that $\sigma^{2}_{min}(\textbf{X}_{A}) = |x_{1}|^2 + |x_{2}|^{2}.$ From this expression, it is straightforward to observe that $\sigma_{min}(\mathcal{C}_{\mathcal{D}_{C}}) > 0$ for $\textbf{X}_{A}$. In Fig. \ref{berl}, we present the bit error rate (BER) of the Alamouti code for the $2 \times 1$ MIMO channel when decoded with (\emph{i}) the IF receiver, and (\emph{ii}) the ML decoder. The plots confirm that the Alamouti code provides full-diversity with the IF receiver. It is well known that Alamouti code is ML decodable with lower computational complexity than the IF receiver. Despite its increased complexity with IF receiver, we have used the IF receiver for Alamouti code only to demonstrate that linear designs with the RNVS property provide full-diversity for the IF linear receiver. 

\begin{figure}
\begin{center}
\includegraphics[scale = 0.6]{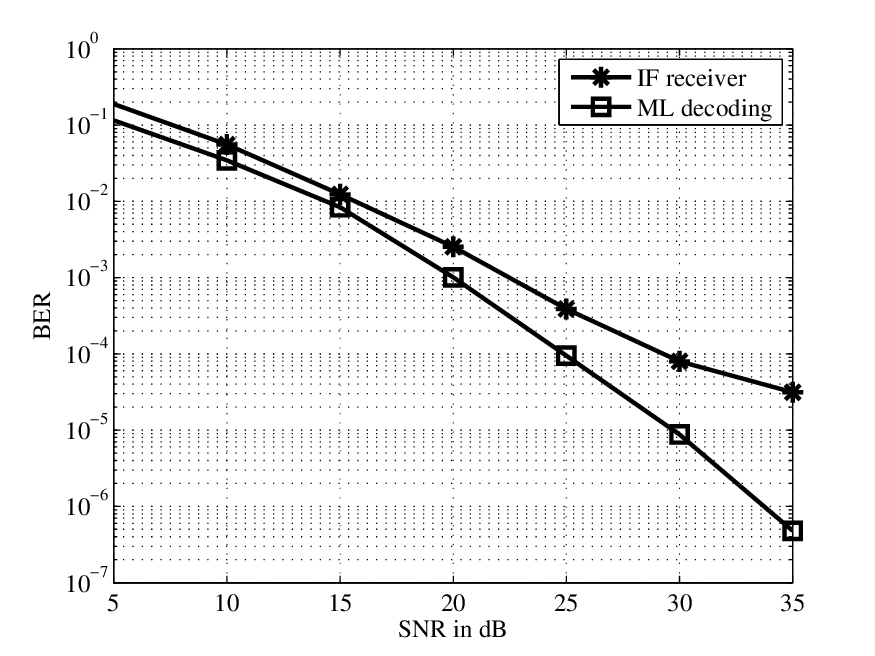}
\caption{\label{ber2}BER comparison of the example code in \eqref{example_code} with IF linear receiver and the ML decoder.}
\end{center}
\end{figure}

We have shown that the NVD property is a sufficient condition to achieve full-diversity. However, we have not shown that it is also a necessary condition. We now present an example of a linear design that satisfies the rank criterion over $\mathcal{S}$ but not the RNVS property. Such a design is given by
\begin{equation}
\label{example_code}
\textbf{X}_{E} = \left[\begin{array}{cc}
x_{1} & 2x_{2}\\
2x_{2} & x_{1}\\
\end{array}\right],
\end{equation}
where $x_{1}, x_{2}\in \mathcal{S}$ carry information symbols. If $\mathcal{S} = \{0, 1, \imath, 1+\imath \},$ then the above linear design satisfies the rank criterion over $\mathcal{S}$, and hence, provides full-diversity for the ML decoder. We can verify that the above linear design does not satisfy the NVD property as $|\mbox{det}(\textbf{X}_{E})|^2 = 0$ for $x_{1} = 4$ and $x_{2} = 2$. In Fig. \ref{ber2}, we present the BER of the above code for the $2 \times 1$ MIMO channel when decoded with (\emph{i}) the IF receiver, (\emph{ii}) the ML decoder. The plots show that this code provides full-diversity with the ML decoder but not with the IF receiver.

\subsection{Relevance of the RNVS property}
\label{subsec1_sec6}

We show through an example that the criterion of maximizing the minimum singular value $\sigma_{min}(\mathcal{C}_{\mathcal{D}_{C}})$ is relevant to design good codes for the IF receiver. We pick two designs of identical symbol-rate that satisfy the RNVS property and show that the one with larger $\sigma_{min}(\mathcal{C}_{\mathcal{D}_{C}})$ performs better than the other. The two designs are given by 
\begin{equation}
\label{example_code_with_nvs}
\textbf{X}_{CE} = \left[\begin{array}{cc}
x_{1} & x_{2}\\
x_{2} & \gamma x_{1}\\
\end{array}\right] \mbox{ and }\textbf{X}_{A} = \left[\begin{array}{cc}
x_{1} & x_{2}\\
-x^{*}_{2} & x^{*}_{1}\\
\end{array}\right],
\end{equation}
where $\textbf{X}_{CE}$ is obtained from the cyclic field extension \cite{BRS} and $\textbf{X}_{A}$ is the Alamouti design obtained from orthogonal designs. For $\textbf{X}_{CE}$, $\gamma = \frac{1}{\sqrt{2}} + \frac{\imath}{\sqrt{2}}$, where $\imath = \sqrt{-1}$. Using the structure of the above designs, it can be shown that 
\begin{equation}
\label{sv_CE}
\sigma^{2}_{min}(\textbf{X}_{CE}) = |x_{1}|^2 + |x_{2}|^{2} - |x_{1}x^{*}_{2} + \gamma^{*}x^{*}_{1}x_{2}|
\end{equation}
and
\begin{equation}
\label{sv_A}
\sigma^{2}_{min}(\textbf{X}_{A}) = |x_{1}|^2 + |x_{2}|^{2}.
\end{equation}
From the above expressions, we first show that $\sigma_{min}(\mathcal{C}_{\mathcal{D}_{C}}) > 0$ for both $\textbf{X}_{CE}$ and $\textbf{X}_{A}$.

\begin{proposition}
Both $\textbf{X}_{CE}$ and $\textbf{X}_{A}$ satisfy the RNVS property.
\end{proposition} 
\begin{proof}
From \eqref{sv_A}, the proof is straightforward for $\textbf{X}_{A}$. Henceforth, we provide the proof for $\textbf{X}_{CE}$. Using the triangle inequality, we have
\begin{equation}
\label{tri_ineq}
|x_{1}x^{*}_{2} + \gamma^{*}x^{*}_{1}x_{2}| \leq |x_{1}x^{*}_{2}| + |\gamma^{*}x^{*}_{1}x_{2}|,
\end{equation}
where the equality holds only when $\frac{x_{1}x^{*}_{2}}{\gamma^{*}x^{*}_{1}x_{2}} = v$ for some constant $v \in \mathbb{R}$. In particular, since $|x_{1}x^{*}_{2}| = |x^{*}_{1}x_{2}|$ and $|\gamma^{*}| = 1$, the equality holds for $v = 1$ or $v = -1$, i.e., when $\frac{x_{1}x^{*}_{2}}{x^{*}_{1}x_{2}} = \pm \gamma^{*}$. However, the constraint $x_{1}, x_{2} \in \mathbb{Z}[\imath]$ guarantees that $\frac{x_{1}x^{*}_{2}}{x^{*}_{1}x_{2}} \in \mathbb{Q}(\imath)$, which in turn implies that $\frac{x_{1}x^{*}_{2}}{x^{*}_{1}x_{2}} \neq \gamma^{*}$ for all $[x_{1} ~x_{2}] \in \mathcal{D}_{C}$. Therefore, the triangle inequality in \eqref{tri_ineq} admits the strict inequality as
\begin{equation}
\label{tri_ineq_strict}
|x_{1}x^{*}_{2} + \gamma^{*}x^{*}_{1}x_{2}| < |x_{1}x^{*}_{2}| + |\gamma^{*}x^{*}_{1}x_{2}|.
\end{equation}
Applying the above bound in \eqref{sv_CE}, we write
\begin{eqnarray}
\sigma^{2}_{min}(\textbf{X}_{CE}) & > & |x_{1}|^2 + |x_{2}|^{2} - |x_{1}x^{*}_{2}| - |\gamma^{*}x^{*}_{1}x_{2}|, \nonumber\\
& \geq & |x_{1}|^2 + |x_{2}|^{2} - |x_{1}||x^{*}_{2}| - |x^{*}_{1}||x_{2}|,\nonumber \\
& = & (|x_{1}| - |x_{2}|)^{2}, \nonumber
\end{eqnarray}
where the second lower bound follows from the Cauchy-–Schwarz inequality and $|\gamma^{*}| = 1$. Thus, $\sigma^{2}_{min}(\textbf{X}_{CE}) > 0$ for all $[x_{1} ~x_{2}] \in \mathcal{D}_{C}$, and hence, the design $\textbf{X}_{CE}$ satisfies the RNVS property.
\end{proof}

\begin{figure}
\begin{center}
\includegraphics[scale = 0.6]{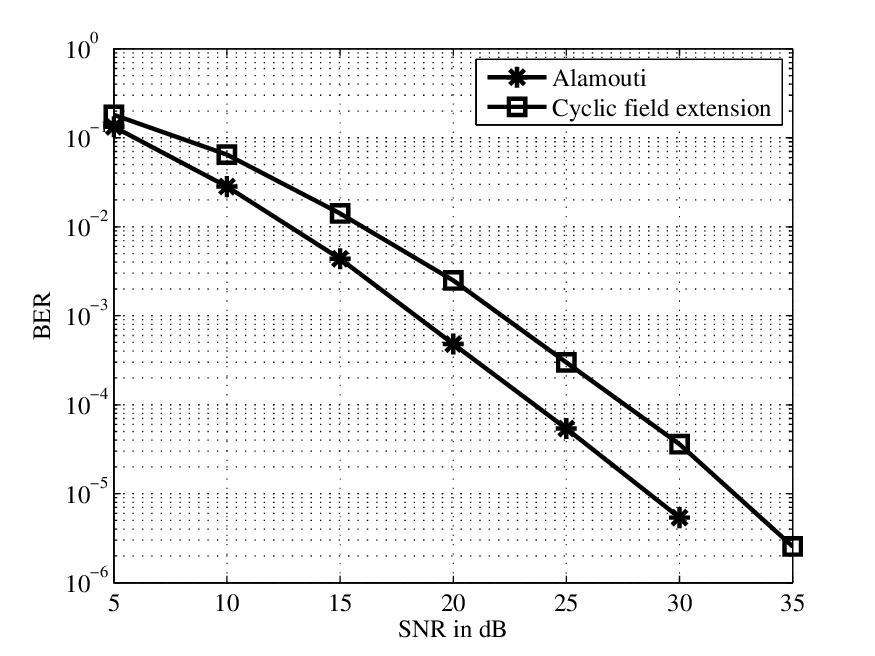}
\caption{\label{ber3}BER comparison of the STBCs in \eqref{example_code_with_nvs} with the IF receiver}
\end{center}
\end{figure}

The above proposition shows that the performance of STBCs from $\textbf{X}_{CE}$ and $\textbf{X}_{A}$ are comparable under IF receiver. From \eqref{sv_CE}, it can be verified that $\sigma_{min}(\textbf{X}_{CE}) < 1$ for $x_{1} = x_{2} = 1$, which in turn implies that $\sigma_{min}(\mathcal{C}_{\mathcal{D}_{C}}) < 1$ for $\textbf{X}_{CE}$. Also, it is straightforward to check that $\sigma_{min}(\mathcal{C}_{\mathcal{D}_{C}}) \geq 1$ for $\textbf{X}_{A}$. Thus, $\textbf{X}_{A}$ is a superior design over $\textbf{X}_{CE}$ with respect to the RNVS property. In Fig. \ref{ber3}, we present the BER of the above designs for the $2 \times 2$ MIMO channel when $\mathcal{S} = \{0, 1, \imath, 1+\imath \}$. The plots in Fig. \ref{ber3} indicate that the order of the BER curves is consistent with the order $\sigma_{min}(\mathcal{C}_{\infty})$ for $\textbf{X}_{A}$ and $\textbf{X}_{CE}$. Thus, we have exemplified the relevance of the RNVS property for the IF receiver. It is important to note that the linear design $\mathbf{X}_{CE}$ satisfies the RNVS property, but not the NVD property, thereby qualifying as a special STBC outside the class of perfect codes that provides full-diversity with IF receivers.

\section{Directions for Future Work}
\label{sec7}

\begin{small}
\begin{table}
\begin{center}
\caption{\label{updated_complexity_table}Comparison of STBCs for various receivers}
\begin{tabular}{|c|c|c|c|} \hline
Approach & Decoding & Symbol- & Equalization\\
& Complexity & Rate & Complexity\\
\hline
{\textbf ML} & high & $\leq \mbox{min}(n_{t}, n_{r})$ & low\\
{\textbf ZF \& MMSE} & low & $\leq 1$ & low\\
{\textbf IF} & low & $\leq \mbox{min}(n_{t}, n_{r})$ & high\\
\hline
\end{tabular}
\end{center}
\end{table}
\end{small}

We have presented a decoder analysis for the IF receiver in order to obtain a design criterion for full-diversity STBCs. We have proposed the restricted non-vanishing singular value (RNVS) property, and have shown that STBCs that satisfy the RNVS criterion provide full-diversity for the IF linear receiver. As a by-product, we have also shown that STBCs with the NVD criterion can be used with the IF receiver. Since perfect codes satisfy the NVD property, our results independently confirm the full-diversity results of perfect codes presented in \cite{OrE}. Importantly, with reference to the question in Fig. \ref{fig:motivation}, we have shown the existence of a code outside the class of perfect codes that provide full-diversity with IF decoding. In summary, among the class of linear receivers, IF receivers admit STBCs with larger spectral efficiency than that of the MMSE and ZF receivers. To conclude, we list down the various properties of the linear receivers in Table \ref{updated_complexity_table}, which shows that the reduction in the decoding complexity for IF receivers comes at the cost of increased complexity in equalization in comparison with the ML decoder. In this context, the term equalization refers to the process of obtaining the matrices $\mathbf{B}$ and $\mathbf{A}$ (as a function of $\mathbf{H}$), which is known to be computationally complex for larger values of $n_{t}$ and $n_{r}$ \cite{Sakzad14-1}. Pointing at this equalization complexity, we highlight that IF linear receivers are suitable for quasi-static fading channels with coherence-time long enough to accommodate multiple space-time codewords; this way the equalization algorithm (such as the one in \cite{HAV1}) is executed once in the coherence-block, and the decoding complexity associated with implementing \textbf{Step 1} to \textbf{Step 3} is that of solving a system of linear equations. An interesting direction for future work is to construct new STBC designs with large value of $\sigma_{min}(\mathcal{C}_{\mathcal{D}_{C}})$ so that they perform well with IF receivers. In this work, although RNVS property has been chosen as a sufficient criterion for full-diversity STBCs, the process of verifying the RNVS property is not straightforward for arbitrary designs. In the case of the $2 \times 2$ STBC from cyclic field extension (presented in Section \ref{subsec1_sec6}), the algebraic structure on the design was used to verify the RNVS property. However, in general, verifying the RNVS property on higher-order designs that do not have any algebraic structure, is not straightforward. On that light, questions related to, how to verify the RNVS property? or how to develop other design-criteria that are easy to verify? are certainly interesting directions for future research.
%
%

\end{document}